\documentstyle[aps,pra]{revtex}
\tightenlines
\title{A two-slit experiment which distinguishes between standard and\\ Bohmian quantum mechanics}
\author{M. Golshani \footnote{E-mail:
golshani@ihcs.ac.ir} and O. Akhavan \footnote{E-mail:
akhavan@mehr.sharif.ac.ir}}
\address{Department of Physics, Sharif University of Technology, P.O. Box 11365--9161, Tehran,  Iran
\\Institute for Studies in Theoretical Physics and Mathematics, P.O. Box 19395--5531, Tehran, Iran}
\begin{document}
\maketitle

\begin{abstract}
In this investigation, we have suggested a special two-slit
experiment which can distinguish between the standard and the
Bohmian quantum mechanics.  At the first step, we have shown that
observable individual predictions obtained from these two
theories are inconsistent for a special case. But, at the ensemble
level, they are consistent as was expected. Then, as another
special case and using selective detection, it is shown that an
observable disagreement between the two theories can exist at the
ensemble level of particles.  This can encourage new efforts for
finding other inconsistencies between the two theories,
theoretically and experimentally.
\end{abstract}

\pacs{ }PACS number: 03.65.Bz \maketitle


\section{Introduction}
Since the standard quantum mechanics (SQM) and the Bohmian quantum
mechanics (BQM) have similar sets of equations, it seems that
these two must be empirically equivalent, particularly at the
ensemble level of particles. Bohm and his collaborators believed
that their theory yields, in every conceivable experiment, the
same statistical results as SQM [1-4]. Bohm, himself, in
responding to the question of whether there is any new prediction
by his theory, said (1986): {\it ``Not the way it's done.  There
are no new predictions because it is a new interpretation to the
same theory"} {\cite{Cus}}.  In fact, when Bohm presented his
theory in 1952, experiments could be done with an almost
continuous beam of particles, but not with individual particles.
Thus, Bohm cooked his theory in such a fashion that it would be
impossible to distinguish his theory from SQM.  For this reason,
when Bell {\cite{Bel}} talked about the empirical equivalence of
the two theories, he was more cautious: {\it ``It [the de
Broglie-Bohm version of non-relativistic quantum mechanics] is
experimentally equivalent to the usual version insofar as the
latter is unambiguous"}. Thus the question arises as to whether
there are phenomena which are well defined in one theory (due to
the presence of path for particles) but ambiguous in the other
one or whether there are experiments which have different
observable results in the two theories? At first, it seems that
the transition of a quantum system through a potential barrier
provides a good case. For this case, there is no well defined
transit time between the two ends of the barrier in SQM, because
time is considered to be a parameter and not a dynamical variable
having a corresponding Hermitian operator. For BQM, however, the
passage of a particle between any two points is conceptually well
defined.  But, the recent work of Abolhasani and Golshani
{\cite{Abo}} indicates that it is not practically feasible to use
this experiment to distinguish between these two theories.
However, there are new efforts to show that BQM can have different
predictions from SQM. For example, Neumaier {\cite{Neu}} claimed
that BQM contradicts SQM.  But, Marchildon {\cite{Mar}} has argued
that this claim is unfounded. Very recently, Ghose {\cite{Gho}}
has claimed that by devising a new version of the two-slit
experiment in which the wave packets of a pair of momentum
correlated identical bosonic particles are simultaneously
diffracted from the two slits, one can distinguish between the
two theories.  Although Ghose's work {\cite{Gho}} is also disputed
by Marchildon {\cite{Mar}}, but Ghose still believes that his
basic conclusions
 are right {\cite{Ghos}}.

In this work, in parallel to Ghose's conclusion in {\cite{Gho}},
we have studied BQM's predictions about a two-slit experiment
with a different particular source to show existence of
disagreement between SQM and BQM for two special cases of the
experiment.

\section{Specifications of the two-slit experiment}

We have considered the following two-slit experiment.  A pair of
identical non-relativistic particles (bosonic or fermionic)
originate simultaneously from a point source.  We assume that the
intensity of the beam is so low that at a time we have only a
single pair of particles passing through the slits. Since the
direction of the emission of each particle can be considered to
be random, we assume that the detection screen registers only
those pairs of particles that reach it simultaneously. Then, the
interference effects of single particles are eliminated on the
screen. Furthermore, it is assumed that the detection process has
no causal role in the phenomenon of interference {\cite{Hol}}.  In
the coordinate system $(x,y)$, the centers of the two slits with
width $2\sigma_0$ are located at $(0,\pm Y)$. We take the
incident wave to be a plane wave of the form
\begin{equation}
\psi_{in}(x_{1},y_{1};x_{2},y_{2};t)=ae^{i[k_{x}(x_{1}+x_{2})+k_{y}(y_{1}+y_{2})]}e^{-iEt/\hbar},
\end{equation}
where $a$ is a constant and
$E=E_{1}+E_{2}=\hbar^{2}(k_{x}^{2}+k_{y}^{2})/m$ is the total
energy of the two-particle system.  We assume that, the source is
located very far from the two-slit screen on the $x$-axis as
compared with $Y$, so that the approximation $k_{y}\simeq 0$ is
valid.  For mathematical simplicity, we avoid slits with sharp
edges which produce mathematical complexity of Fresnel
diffraction, i.e., we assume that the slits have soft edges, so
that the Gaussian wave packets are produced along the
$y$-direction, and that the plane waves along the $x$-axis remain
unchanged {\cite{Hol}}.  In fact, the one-particle wave function
should be represented by Gaussian wave packets rather than plane
or spherical waves, as utilized by Ghose {\cite{Gho}} and
Marchildon {\cite{Mar}}, respectively. We take the time of the
formation of the Gaussian wave to be $t=0$. Then, the emerging
wave packets from the slits $A$ and $B$ are
\begin{equation}
\psi_{A}(x,y)=a(2\pi\sigma_{0}^{2})^{-1/4}e^{-(y-Y)^{2}/4\sigma_{0}^{2}}e^{i[k_{x}x+k_{y}(y-Y)]},
\end{equation}
\begin{equation}
\psi_{B}(x,y)=a(2\pi\sigma_{0}^{2})^{-1/4}e^{-(y+Y)^{2}/4\sigma_{0}^{2}}e^{i[k_{x}x-k_{y}(y+Y)]},
\end{equation}
where $\sigma_{0}$ is the half-width of each slit.

 Now, for this
two-particle system, the total wave function at the detection
screen, at time $t$, is
\begin{eqnarray}
\psi(x_{1},y_{1};x_{2},y_{2};t)=&N&[\psi_{A}(x_{1},y_{1},t)\psi_{B}(x_{2},y_{2},t)+\psi_{A}(x_{2},y_{2},t)\psi_{B}(x_{1},y_{1},t)\nonumber\\
 &+&\psi_{A} (x_{1},y_{1},t)\psi_{A}(x_{2},y_{2},t)+\psi_{B}
(x_{1},y_{1},t)\psi_{B} (x_{2},y_{2},t)]
\end{eqnarray}
with
\begin{equation}
\psi_{A}(x,y,t)=a(2\pi
\sigma_{t}^{2})^{-1/4}e^{-(y-Y-u_{y}t)^{2}/4\sigma_{0}\sigma_{t}}e^{i[k_{x}x+k_{y}(y-Y-u_{y}t/2)-E_{x}t/\hbar]},
\end{equation}
\begin{equation}
\psi_{B}(x,y,t)=a(2\pi
\sigma_{t}^{2})^{-1/4}e^{-(y+Y+u_{y}t)^{2}/4\sigma_{0}\sigma_{t}}e^{i[k_{x}x-k_{y}(y+Y+u_{y}t/2)-E_{x}t/\hbar]},
\end{equation}
where $N=1/[2(1+e^{(-Y^{2}/2\sigma_{0}^{2})})]$ is a
reparameterization constant,
\begin{equation}
\sigma_{t}=\sigma_{0}(1+\frac{i\hbar t}{2m\sigma_{0}^{2}})
\end{equation}
and
\begin{eqnarray}
&& u_{y}=\frac{\hbar k_{y}}{m},\cr&&
E_{x}=\frac{1}{2}mu_{x}^{2},
\end{eqnarray}
where $u_{x}$ and $u_{y}$ are initial group velocities
corresponding to each particle in the $x$ and $y$-directions,
respectively.

It is well known from SQM that the probability of simultaneous
detection of the particles at $y_1=Q_{1}$ and $y_2=Q_{2}$, on the
screen , located at $x_{1}=x_{2}=D$ at $t=D/u_{x}$, is equal to
\begin{equation}
P_{12}(Q_{1},Q_{2},t)=\int_{Q_{1}}^{Q_{1}+\triangle}dy_{1}\int_{Q_{2}}^{Q_{2}+\triangle}dy_{2}|\psi(x_{1},y_{1};x_{2},y_{2};t)|^{2}.
\end{equation}
The parameter $\Delta$, which is taken to be small,  is a measure
of the size of the detectors.  We shall compare this prediction
of SQM with that of BQM.

\section{The predictions of BQM for the suggested experiment}

In BQM,  the complete description of a system is given by
specifying the location of its particles, in addition to the wave
function which has the role of guiding the particles according to
the following guidance condition:
\begin{equation}
\overrightarrow{\dot{x}_{i}}(\overrightarrow{x},t)=\frac{1}{m_{i}}
\overrightarrow{\nabla_{i}}S(\overrightarrow{x},t)=\frac{\hbar}{m_{i}}Im\left(
\frac{\overrightarrow{\nabla_{i}}\psi(\overrightarrow{x},t)}{\psi(\overrightarrow{x},t)}\right),
\end{equation}
where $S(\overrightarrow{x},t)$ is the phase of total wave
function which is written in polar form
\begin{equation}
\psi(\overrightarrow{x_1},\overrightarrow{x_2},...,\overrightarrow{x_n};t)=
R(\overrightarrow{x_1},\overrightarrow{x_2},...,\overrightarrow{x_n};t)
e^{iS(\overrightarrow{x_1},\overrightarrow{x_2},...,\overrightarrow{x_n};t)/\hbar}.
\end{equation}
 Here, the speed of the particles 1 and 2 in the direction $y$
is given, respectively, by
\begin{equation}
\dot{y}_{1}(x_{1},y_{1};x_{2},y_{2};t)=\frac{\hbar}{m}Im\frac{\partial_{y_{1}}\psi(x_{1},y_{1};x_{2},y_{2};t)
}{\psi(x_{1},y_{1};x_{2},y_{2};t)},
\end{equation}
\begin{equation}
\dot{y}_{2}(x_{1},y_{1};x_{2},y_{2};t)=\frac{\hbar}{m}Im\frac{\partial_{y_{2}}\psi(x_{1},y_{1};x_{2},y_{2};t)
}{\psi(x_{1},y_{1};x_{2},y_{2};t)}.
\end{equation}
With the replacement of $\psi(x_{1},y_{1};x_{2},y_{2};t)$ from
(4), we have
\begin{eqnarray}
\dot{y}_{1}=&N&\frac{\hbar}{m}Im\{\frac{1}{\psi}[[-2(y_{1}-Y-u_{y}t)/4\sigma_{0}\sigma_{t}+ik_{y}]\psi_{A_{1}}\psi_{B_{2}}\cr
           &+&[-2(y_{1}+Y+u_{y}t)/4\sigma_{0}\sigma_{t}-ik_{y}]\psi_{A_{2}}\psi_{B_{1}}\cr
           &+&[-2(y_{1}-Y-u_{y}t)/4\sigma_{0}\sigma_{t}+ik_{y}]\psi_{A_{1}}\psi_{A_{2}}\cr
           &+&[-2(y_{1}+Y+u_{y}t)/4\sigma_{0}\sigma_{t}-ik_{y}]\psi_{B_{1}}\psi_{B_{2}}]\},
\end{eqnarray}
\begin{eqnarray}
\dot{y}_{2}=&N&\frac{\hbar}{m}Im\{\frac{1}{\psi}[[-2(y_{2}+Y+u_{y}t)/4\sigma_{0}\sigma_{t}-ik_{y}]\psi_{A_{1}}\psi_{B_{2}}\cr
           &+&[-2(y_{2}-Y-u_{y}t)/4\sigma_{0}\sigma_{t}+ik_{y}]\psi_{A_{2}}\psi_{B_{1}}\cr
           &+&[-2(y_{2}-Y-u_{y}t)/4\sigma_{0}\sigma_{t}+ik_{y}]\psi_{A_{1}}\psi_{A_{2}}\cr
           &+&[-2(y_{2}+Y+u_{y}t)/4\sigma_{0}\sigma_{t}-ik_{y}]\psi_{B_{1}}\psi_{B_{2}}]\}.
\end{eqnarray}
On the other hand, from (5) and (6) one can see that,
\begin{eqnarray}
&&\psi_{A}(x_{1},y_{1},t)=\psi_{B}(x_{1},-y_{1},t),\cr
&&\psi_{A}(x_{2},y_{2},t)=\psi_{B}(x_{2},-y_{2},t),
\end{eqnarray}
 which indicates the reflection symmetry  of $\psi(x_{1},y_{1};x_{2},y_{2};t)$ with
respect to the $x$-axis. Using this symmetry in (14) and (15), we
have
\begin{eqnarray}
&&\dot{y}_{1}(x_{1},y_{1},t)=-\dot{y}_{1}(x_{1},-y_{1},t),\cr&&
\dot{y}_{2}(x_{2},y_{2},t)=-\dot{y}_{2}(x_{2},-y_{2},t).
\end{eqnarray}
 These relations show that, if $y_{1}(t)=0$ or $y_{2}(t)=0$, then the speed
of each particles in the $y$-direction is zero along the symmetry
axis $x$.  This means that, none of the particles can cross the
$x$-axis nor are tangent to it.  The fact that the paths of the
two particles are located on the two sides of the $x$-axis can
lead, under suitable conditions, to a discrepancy between the
predictions of SQM and BQM, particularly at the ensemble level.

If we consider $y=(y_{1}+y_{2})/2$ to be the vertical coordinate
of the center of mass of the two particles, then we can write
\begin{eqnarray}
\dot{y}&=&(\dot{y}_{1}+\dot{y}_{2})/2\cr
        &=&N\frac{\hbar}{2m}Im\{\frac{1}{\psi}[(-\frac{y_{1}+y_{2}}{2\sigma_{0}\sigma_{t}})(\psi_{A_{1}}\psi_{B_{2}}+\psi_{A_{2}}\psi_{B_{1}}+\psi_{A_{1}}\psi_{A_{2}}+\psi_{B_{1}}\psi_{B_{2}})\cr
       &&+(\frac{Y+u_{y}t}{\sigma_{0}\sigma_{t}}+2ik_{y})(\psi_{A_{1}}\psi_{A_{2}}-\psi_{B_{1}}\psi_{B_{2}})]\}\cr
       &=&\frac{(\hbar/2m\sigma_{0}^{2})^{2}t(y_{1}+y_{2})/2}{1+(\hbar/2m\sigma_{0}^{2})^{2}t^{2}}+N\frac{\hbar}{2m}Im\{\frac{1}{\psi}(\frac{Y+u_{y}t}{\sigma_{0}\sigma_{t}}+2ik_{y})(\psi_{A_{1}}\psi_{A_{2}}-\psi_{B_{1}}\psi_{B_{2}})\}.
\end{eqnarray}
Now, we consider the two following special cases:\\
\subsection{The $\langle y_0\rangle=0$ and $\hbar t/2m\sigma_{0}^{2}\sim 1$ condition}
At first, consider the conditions $k_{y}\simeq 0$ and
$Y\ll\sigma_0$.  Then, we can consider
\begin{equation}
\psi_{A_{1}}\psi_{A_{2}}-\psi_{B_{1}}\psi_{B_{2}}\simeq0,
\end{equation}
in the second term of eq. (18).  Hence, that equation of motion
for the $y$-coordinate of the center of mass is converted to
\begin{equation}
\dot{y}\simeq\frac{(\hbar/2m\sigma_{0}^{2})^{2}}{1+(\hbar/2m\sigma_{0}^{2})^{2}t^{2}}yt.
\end{equation}
Had we omitted the last two terms in the wave function (4), as was
done in {\cite{Gho}}, we would have obtained the same  result,
exactly. Solving the differential equation (20), we get the path
of the $y$-coordinate  of the center of mass in the form
\begin{equation}
y\simeq y_{0}\sqrt{1+(\hbar/2m\sigma_{0}^{2})^{2}t^{2}}.
\end{equation}
If the center of mass of the system was exactly on the $x$-axis
at $t=0$, i.e. $y_{0}=0$, then the center of mass of the system
would have always remained on the $x$-axis and all the joint
detections of the two particles would have happened symmetrically
around the $x$-axis.  However, based on the quantum equilibrium
hypothesis, $y_0$ has an initial quantum distribution according
to $|\psi|^2$ at $t=0$.  Thus, it may seem that we do not have
symmetrical detection around the $x$-axis.  On the other hand, it
is well known that, the distance between any two neighboring
maxima is nearly given by $\delta y\simeq\lambda x/2Y$, where
$\lambda$ is the de Broglie wavelength.  If we consider $\delta
y$ as a length scale on the screen, then to obtain a symmetrical
detection with a reasonable approximation, it would be enough
that the center of mass deviation from the $x$-axis to be
considered smaller than the distance between the neighboring
maxima, that is,
\begin{equation}
\triangle y\ll\delta y\simeq\frac{\lambda
D}{2Y}\simeq\frac{\pi\hbar t }{Ym}.
\end{equation}
By considering the conditions $\hbar t/2m\sigma_{0}^{2}\sim 1$ and
$\triangle y_0\sim\sigma_0$, one can obtain that the constraint
of symmetrical detection of the two particles is
\begin{equation}
Y\ll 2\pi\sigma_0,
\end{equation}
which is consistent with our applied initial assumption on $Y$ and
$\sigma_0$ in eq. (19).  But, in SQM, there is a non-zero
probability to find the two particles asymmetrical on the screen,
as can be seen by eq. (9).  Therefore, we have obtained a
disagreement between the asymmetrical prediction of SQM and the
symmetrical prediction of BQM for our special conditions.

It is worthy to note that, since the wave function specified in
eq. (4) can be written in a factorizable form of wave function,
namely
\begin{equation}
\psi(x_1,x_2;y_1,y_2;t)=N[\psi_A(x_1,y_1,t)\psi_B(x_2,y_2,t)][\psi_A(x_2,y_2,t)\psi_B(x_1,y_1,t)],
\end{equation}
it can be seemed that the two particles behave independently at
the ensemble level.  Thus, SQM and BQM predict the same
interference pattern for an ensemble pair of particles, as
expected.
\subsection{The $\langle y_0\rangle\neq 0$ and $\hbar t/2m\sigma_{0}^{2}\gg 1$ condition}
In this case, we once again use eq. (21) for describing the time
development of the center of mass $y$-coordinate, under the
conditions $k_{y}\simeq 0$ and $Y\ll\sigma_0$.  Furthermore, we
try to perform our experiment in the following fashion: we record
only those particles which are detected on the two sides of the
$x$-axis, simultaneously.  That is, we eliminate the cases of
detecting only one particle or detecting the pairs which pass
through the same slit, which means that we apply a selective
detection on the particles.  Thus, based on BQM, particularly the
results obtained from (17), there will be a length
\begin{equation}
L\simeq2\langle y\rangle\simeq\frac{\hbar t\langle
y_{0}\rangle}{m\sigma_{0}^{2}}
\end{equation}
 on the screen where almost no particle is recorded under the condition $\hbar t/2m\sigma_{0}^{2}\gg 1$,
  if the constraint $\triangle y\ll L$ is satisfied.  This constraint will turn into
\begin{equation}
\sigma_0\ll\langle y_0\rangle,
\end{equation}
if the quantum equilibrium constraint $\triangle y_0\sim\sigma_0$
is considered.

However, based on SQM, we have two alternatives:\\
i) The probability relation (9) is still valid and due to the
selective detection there is only
a reduction in the intensity of the particles.\\
ii) SQM is silent about our selective detection.\\
In the first case, there is a disagreement between predictions of
SQM and BQM and in the second case, BQM shows a better predictive
power than SQM, even at the ensemble level. Therefore, it seems
that performing such experiment provides observable differences
between the two theories, particularly at the ensemble level.

\section{Conclusion}
We have shown that, a two-slit experiment, with a special source
emitting two unentangled identical particles and with the
condition $Y\ll\sigma_0$, can yield different predictions for SQM
and BQM, in two special cases. In the case $\langle y_0\rangle=0$
and $\hbar t/2m\sigma_{0}^{2}\sim 1$, BQM results symmetrical
detection of the two particles around the $x$-axis with
reasonable approximation, while according to SQM the probability
of finding the two particles at arbitrary points on the screen is
not zero. Furthermore, in the case $\langle y_0\rangle\neq 0$ and
$\hbar t/2m\sigma_{0}^{2}\gg 1$, BQM predicts an empty interval on
the screen if one uses selective detection, which is not
predictable by SQM. Therefore, this experiment seems to shed
light on the question of whether wave function provides a
complete description of a system.

\end{document}